\documentclass[a4paper, amsfonts, amssymb, amsmath, reprint, showkeys, nofootinbib, twocolumn, twoside, aps]{revtex4}
\usepackage[english]{babel}
\usepackage[utf8]{inputenc}
\usepackage[colorinlistoftodos, color=green!40, prependcaption]{todonotes}
\usepackage{amsthm}
\usepackage{mathtools}
\usepackage{xcolor}
\usepackage{graphicx}
\usepackage[left=23mm,right=13mm,top=35mm,columnsep=15pt]{geometry} 
\usepackage{adjustbox}
\usepackage[T1]{fontenc}
\usepackage{lipsum}
\usepackage{csquotes}
\usepackage{afterpage}
\usepackage{placeins}
\usepackage{subfigure}
\usepackage{float}
\usepackage{tikz}
\usepackage{tikz-feynman}
\tikzfeynmanset{compat=1.1.0}
\usepackage[pdftex, pdftitle={Article}, pdfauthor={Author}]{hyperref} 
\begin{document}

\begin{flushright}
\begin{small}
\end{small}
\end{flushright}
\title{Bounds on decaying sterile neutrinos via magnetic dipole moment from COB intensity}

\author{Hriditi Howlader}
\email{p22ph005@nitm.ac.in}
 \affiliation{Department of Physics, National Institute of Technology Meghalaya, Shillong, Meghalaya, India}

 \author{Vivekanand Mohapatra}
 \email{p22ph003@nitm.ac.in}
 \affiliation{Department of Physics, National Institute of Technology Meghalaya, Shillong, Meghalaya, India}

\author{Alekha C. Nayak}
    \email{alekhanayak@nitm.ac.in}
    \affiliation{Department of Physics, National Institute of Technology Meghalaya, Shillong, Meghalaya, India}

\author{Tripurari Srivastava}
    \email{tripurarisri022@gmail.com}
    \affiliation{Institute of Particle Physics and Key Laboratory of Quark and Lepton Physics (MOE),
Central China Normal University, Wuhan, Hubei 430079, China}
    \affiliation{Department of Theoretical Physics, Tata Institute of Fundamental Research,
Homi Bhabha Road, Mumbai 400005, India}


\begin{abstract}

A recent observation by Long
Range Reconnaissance Imager (LORRI) mounted on NASA's New Horizons yielded the most accurate measurement of the cosmic optical background (COB). The reported COB intensity is $11.16\pm 1.35$ $\mathrm{nW/m^2/sr}$ at a pivot wavelength $ \lambda_{piv} = 0.608 \, \mu\mathrm{m}$ observed in the range \( 0.4 \, \mu\mathrm{m} \lesssim \lambda \lesssim 0.9 \, \mu\mathrm{m} \). After subtracting the measured intensity from the deep Hubble Space Telescope count, diffused galactic light, and scattered light from bright star foregrounds, an anomalous intensity of $2.99 \pm 2.03~\mathrm{nW/m^2/sr}$ has been found. 
We considered radiatively decaying sterile neutrinos of keV mass scale, as dark matter candidate, that could contribute to this anomalous reported intensity. Using this, we derive upper bounds on the sterile-to-sterile transition magnetic moment. We find that sterile neutrinos with mass of $\mathcal{O}(\rm keV)$ scale take values of the transition magnetic moment in the range $ 3\times 10^{-13}\,\rm eV^{-1} - 10^{-9}\,\rm eV^{-1}$ to explain the anomalous intensity of $2.99\pm 2.03\,\rm nW/m^2/sr$. 
\end{abstract}
\keywords{ Sterile-to-sterile transition magnetic moment, Low Energy Effective Field Theory (LEFT)}

\maketitle

\section{Introduction}

Line-intensity mapping (LIM) is an observational method for extragalactic astronomy and cosmology that measures the integrated intensity of spectral lines emitted from galaxies and the intergalactic medium at a specific observed frequency \cite{Kovetz:2017agg, Kovetz:2019uss}. Since LIM experiments capture information from all incoming photons, they have the potential to detect electromagnetic radiation produced by dark matter decays directly. For instance, in Ref. \cite{Grin:2006aw}, authors have derived a stringent upper bound on the axion-photon coupling strength by searching for optical line emission produced from decaying axions. In Ref. \cite{Gong:2015hke}, authors have shown that photons produced from decaying axions with eV mass can explain the anisotropy of near-IR extragalactic background light. In Ref. \cite{Creque-Sarbinowski:2018ebl}, the authors have proposed that LIM experiments can detect photons produced from decaying and annihilating dark matter particles. However, measuring the extragalactic background light in the optical band is a difficult task. The primary reason is mitigating the overwhelming foreground, which requires accurate modelling and well-calibrated instruments \cite{bernal2022cosmic}. In Ref. \cite{zemcov2017measurement}, authors have suggested observing such radiations in the optical spectrum—the cosmic optical background (COB) is possible if one can eliminate the foreground radiations, primarily solar in nature. 

The Long Range Reconnaissance Imager (LORRI) mounted on NASA New Horizons spacecraft, currently operating at a distance $51.3$ astronomical unit (AU) from the Sun, has been measuring the COB photon intensity in the wavelength range $0.4\,\mu \mathrm{m}\lesssim\lambda\lesssim 0.9\,\mu \mathrm{m}$ $(\sim 1.3 ~\rm eV - 3 ~\,\rm eV)$ \cite{Zemcov:2017dwy, Lauer:2020qwk, Lauer:2022fgc, Symons:2022lke, Postman:2024erl}. In $2021$, LORRI operating at 42-45 AU from the Sun reported an optical intensity from unknown sources of $8.8\pm 4.9\, \rm nW/m^2/sr$ - $11.9\pm 4.6\, \rm nW/m^2/sr$ \cite{Lauer:2020qwk}. Later, it yielded an anomalous intensity of COB, $8.06\pm 1.92\,\rm nW/m^2/sr$ while operating at $51.3$ AU \cite{Lauer:2022fgc}. In Ref.~\cite{Lauer:2022fgc}, the authors considered diffused galactic light (DGL) from the Milky Way and scattered light from bright stars (SSL) contributing to the total sky level. Further, they subtracted the intensity from integrated light of external galaxies (IGL), known as deep Hubble galaxy count, $8.17 \pm 1.18 \,\rm nW/m^2/sr$ from the total COB level to report the anomalous intensity \cite{Lauer:2022fgc}. Later, this result was also supported by an independent research following blind methodology \cite{Symons:2022lke}. However, in a recent analysis \cite{Postman:2024erl}, the authors have reworked the analysis of the foregrounds from the DGL and found that the contribution of DGL to the COB intensity is greater compared to the one found in the article \cite{Lauer:2022fgc}. Thus yielding an anomalous intensity $2.99 \pm 2.03\, \rm{nW/m^2/sr}$ in the optical spectrum, relatively modest to the IGL intensity \cite{Postman:2024erl}.


 
This anomalous photon excess may indicate the presence of physics beyond the Standard Model.
Various studies have explored such possibilities, for e.g., photons emitted from decaying axions with masses in the range of $\sim 5 ~\rm eV-25~\rm eV$ \cite{bernal2022cosmic, Nakayama:2022jza, Carenza:2023qxh}. Additionally, decaying sterile neutrinos could produce a similar photon signal through their decay into photons and neutrinos. 

Short-baseline neutrino experiments have hinted at the existence of additional neutrinos, and numerous studies have investigated sterile neutrinos as a potential dark matter candidate \cite{ho2013sterile}. The investigation of photon radiated from the sterile neutrinos can provide an exciting probe for the beyond standard model containing sterile neutrinos. Motivated by these possibilities, we analyse photon flux originating from decaying sterile neutrinos. We also explore it as the source of the observed anomalous intensity. 

A sizable transition magnetic moment could enhance photon production from sterile neutrinos, making this anomaly a valuable probe of new physics in the neutrino sector. The transition magnetic moment of sterile neutrino allows it to interact with standard model particles such as active neutrinos, gauge bosons, and photons \cite{ge2023disentangle}. In particle physics, the transition magnetic moment denotes that a sterile neutrino possesses a non-zero intrinsic magnetic orientation, allowing it to interact with external magnetic fields \cite{ge2023unique}. There have been several studies on the phenomenological implications of active-to-active and active-to-sterile neutrino transition magnetic moments, particularly for small neutrino masses, using both effective field theory approaches and models beyond the Standard Model \cite{voloshin1987compatibility,barbieri1989neutrino,babu2020large,babu2021muon,schwetz2021constraining}. Many experiments have placed upper limits on these magnetic moments, with constraints on sterile-to-sterile transition magnetic moment being relatively weaker, especially at smaller mass scales. The sterile-to-sterile transition magnetic moment for neutrinos in the GeV mass range has been studied in the context of long-lived particle (LLP) searches in various experiments, such as AL3X, ANUBIS, CODEX-b, FACET, MAPP, and FASER/FASER$\nu$ \cite{barducci2023probing, gunther2024long}. This research opens new opportunities for LLP searches at the Large Hadron Collider (LHC) \cite{beltran2024probing}. Additionally, measurements from COB provide a complementary method for constraining transition magnetic moments.

In this work, we focus on the keV mass scale of sterile neutrinos, as they are considered a viable dark matter candidate. Rather than considering a specific model, we adopt an effective field theory approach \cite{beltran2024probing}, which makes our analysis more general and less dependent on particular model assumptions. We investigate the potential role of the keV-mass scale sterile neutrinos in contributing to the reported anomalous COB intensity, thus deriving upper bounds on the sterile-to-sterile transition magnetic  moment  by varying the $\Delta m$ values from $2\,\rm eV$ to $10\,\rm eV$. Future experiments like the SPHEREx telescope surveying the sky at wavelengths $0.75-5\, \mu m$ might shed light on the properties of radiatively decaying sterile neutrinos \cite{2020SPIE11443E..0IC, SPHEREx:2014bgr, SPHEREx:2016vbo, SPHEREx:2018xfm}.

The paper is organised as follows. In Section \ref{sec2}, we discuss the radiative decay process of sterile neutrinos through the sterile-to-sterile transition magnetic dipole moment in the low-energy effective field theory. In Section \ref{sec3}, we derive the mean specific intensity for the keV mass range of decaying sterile neutrinos. In Section \ref{sec4}, we have analyzed our results by determining the minimal required sterile-to-sterile transition magnetic moment and decay width of keV-mass scale sterile neutrinos. In this section, we consider the two COB measurements; one is anomalous COB excess intensity, $8.06\pm 1.92\,\rm nW/m^2/sr$ and another one is the rest intensity of COB, $2.99\pm 2.03 \,\rm nW/m^2/sr$. Here, we also derive the novel constraints on the sterile-to-sterile transition magnetic moment through the contribution of the COB intesnity from the optical to X-rays region. In section\ref{sec:uv}, we explain the UV origin of the small mass splitting of sterile neutrinos. In Section \ref{sec:Conclusion}, we present our conclusions with an outlook for future works.

\section{Radiative Decay of Sterile Neutrino through Transition Magnetic  Moment}
\label{sec2}

We consider Standard Model augmented with sterile neutrinos in the effective field theory approach. Assuming sterile neutrino mass scales well below the electroweak scale, we can write the effective Lagrangian density that contributes to the magnetic moment of the neutrino sector: 
\begin{equation}
\label{eq:8}
        \mathcal{L}_{N_{R}LEFT} \supset  d_{\rm NN\gamma}^{ij}\mathcal{O}_{\rm NN\gamma}^{ij} + d_{\nu N\gamma}^{\beta i}\mathcal{O}_{\nu N\gamma}^{\beta i} + \text{h.c.}\,,
\end{equation}
where, $d_{\rm NN\gamma}$ represents the sterile-to-sterile transition magnetic  moment and $d_{\rm \nu N \gamma}$ denotes the active-to-sterile transition magnetic moment, respectively. The term $\mathcal{O}_{\rm NN\gamma}^{ij}=(\bar N_{Ri}^{c}\sigma_{\mu\nu}N_{Rj})F^{\mu\nu}$, $\mathcal{O}_{\nu N\gamma}^{\alpha i}=(\bar \nu_{L\alpha}^{c}\sigma_{\mu\nu}N_{Ri})F^{\mu\nu}$ represent the effective field operators, where $F^{\mu\nu}$ is the electromagnetic field strength tensor. The electric and magnetic field off-diagonal non-zero transition dipole moment components give rise to radiative decay through two different sterile states. Therefore, in this radiative decay process ($m_{1}\rightarrow m_{2}+\gamma$), the decay width of sterile neutrino can be written as \cite{atre2009search,bondarenko2018phenomenology,balantekin2014magnetic}
\begin{equation}
\label{eq: dec width}
    \Gamma_{m_{1}}(m_{1}\rightarrow m_{2}+\gamma) = \frac{2|d_{\rm NN\gamma}|^{2}}{\pi} m_{1}^{3}\left(1-\frac{m_{2}^{2}}{m_{1}^{2}}\right)^{3}.
\end{equation}
By defining $\delta = 1 - \frac{m_{2}}{m_{1}}$, which represents the mass splitting between two eigenstates of sterile neutrino, we can rewrite Eq. \eqref{eq: dec width} as

\begin{equation}
\label{eq: decay width}
    \Gamma_{m_{1}} (m_{1}\rightarrow m_{2}+\gamma)= \frac{2|d_{\rm NN\gamma}|^{2}}{\pi} m_{1}^{3}\left(2-\delta\right)^{3}\delta^{3},
\end{equation}
 From Eq. \eqref{eq: decay width}, we can derive the sterile-to-sterile transition magnetic moment $d_{\rm NN\gamma}$. 

There are several experimental upper bounds on the neutrino magnetic moment. Recent studies of the NOMAND neutrino detector at CERN put constraints on the active-to-sterile transition magnetic moment $d_{\rm \nu N\gamma}\lesssim (10^{-6}-10^{-9})\mu_{B}$ \cite{gninenko1999limits}. Here, $\mu_B\sim 3\times 10^{-7}\,\rm eV^{-1}$ represents the Bohr magneton. In an anti-neutrino electron scattering experiment, by examining the electron recoil spectra, it was found that the active-to-sterile transition magnetic moment $d_{\rm \nu N\gamma}\lesssim 10^{-9}\mu_{B}$ \cite{derbin1993experiment}. The Super-Kamiokande experiment put an upper bound limit on active-to-sterile transition magnetic moment $d_{\rm \nu N\gamma}\lesssim 1.1\times 10^{-10}\mu_{B}$ \cite{liu2004limits}, and the Borexino collaboration put the upper bound limit at $90\%$ confidence level as $5.4\times 10^{-11}\mu_{B}$ \cite{agostini2017limiting,arpesella2008direct}. MUNU collaboration has found the upper bound on active-to-sterile transition magnetic moment $d_{\rm \nu N\gamma}\lesssim 9.0\times 10^{-11}\mu_{B}$ \cite{babu2020large}, whereas TEXONO collaboration has obtained this active-to-sterile transition magnetic  moment upper bound limit $d_{\rm \nu N\gamma}\lesssim 7.4\times 10^{-11}\mu_{B}$ \cite{balantekin2014magnetic}. The Planck+BAO put the constraint on the active-to-sterile transition magnetic moment, and it varies from $3.7\times 10^{-11}\mu_{B}$ to $9.1\times 10^{-12}\mu_{B}$  \cite{carenza2024strong}. Recent studies analysing the cooling of red giant stars have found an upper limit on active-to-sterile transition magnetic moments $d_{\rm  \nu N\gamma} \lesssim 1.2 \times 10^{-12}\mu_{B}$ \cite{capozzi2020axion}. The anomalous stellar cooling due to plasmon decay has been found an upper limit on the active-to-sterile transition magnetic moment $d_{\rm \nu N\gamma}$ is $d_{\rm \nu N\gamma} \lesssim 3 \times 10^{-12} \mu_{B}$ \cite{alexander2016status}.

\section{Mean specific intensity of decaying sterile neutrinos}
\label{sec3}

Consider a radiative decaying particle of rest mass $m_{\chi}$ represented as $\chi\to \gamma+\xi$, where $\gamma$ and $\xi$ represent a photon and another particle, respectively. The wavelength of the photon produced in the rest frame of $\chi$ is \cite{Chen:2003gz, Bernal:2020lkd}
\begin{equation}
    \lambda_e = \frac{hc}{m_{\chi}c^2}\times \frac{2}{\left(1-m_{\xi}^2/m_{\chi}^2\right)}
    \label{eq:lambda_e}
\end{equation}
where $h$ and $c$ are Planck's constant and the speed of light in vacuum, respectively. We define the second term on the right-hand side of Eq. \eqref{eq:lambda_e} as $x^{-1}=\frac{2}{\left(1-m_{\xi}^2/m_{\chi}^2\right)}$. We note that for a two-photon decay process, $x=1/2$. The wavelength of photon produced from a particle decaying at redshift $z$ will be observed today as $\lambda_{obs} = (1+z)\lambda_e$. If the intergalactic gas does not absorb these photons along the line of sight, then they will be observed today as the extragalactic background light. The redshift dependence of the energy density of decaying dark matter particles is given by $\rho_\chi(t)=\rho_\chi(0)(1+z)^3\mathrm{e}^{-\Gamma_{\chi} t}$, where $\Gamma_{\chi}$ represents the decay width, $t$ is the time variable, and $\rho_\chi(0)$ is the dark matter energy density today. The specific intensity of the photons $I_{\nu}$ observed at frequency $\nu_{\rm obs}$ today is given by \cite{Creque-Sarbinowski:2018ebl}

\begin{equation}
    I_\nu = \dfrac{1}{4\pi}\int_0^{\infty}{dz}\dfrac{c}{H(z)}\dfrac{\epsilon_{\nu_{\rm obs}}(z)}{(1+z)^4},
    \label{eq:I_nu}
\end{equation}
where $\epsilon_{\nu_{\rm obs}}$ is the specific luminosity density, and $H(z)$ is the Hubble parameter. For a Dirac delta decaying profile,  $\epsilon_{\nu_{\rm obs}}$ is expressed as \cite{Chen:2003gz, Creque-Sarbinowski:2018ebl}
\begin{equation}
    \epsilon_{\nu_{\rm obs}}(z) = h \mathrm{N}_\gamma\Gamma_\chi\mathrm{n}_\chi(z)\mathrm{E}_\gamma\,\delta(\rm E-E_\gamma),
\end{equation}
where, $\rm N_{\gamma}$ and $\rm n_{\chi}(z)$ represent the number of photons produced and the number density of the decaying particles. On solving Eq. \eqref{eq:I_nu} for the aforementioned $\epsilon_{\nu_{\rm obs}}$ and using the relation, $I_\lambda=I_\nu\left|\mathrm{d\nu}/\mathrm{d\lambda}\right|$ we calculate the mean specific intensity per observed wavelength as 
\begin{equation}
    I_{\lambda} = \frac{c}{4\pi}\frac{\Omega_{\chi,0}\rho_c\Gamma_{\chi}}{\lambda_{\rm obs}(1+z)\text{H}(z)m_{\chi}}~\frac{hc}{\lambda_e}N_{\gamma},
    \label{eq:I_lambda}
\end{equation}
where $\Omega_{\chi,0}$ represents the present-day dark matter density parameter and $\rho_c$ is the critical density. 
In this work, we consider a decaying sterile neutrino of mass $m_1$ to produce a photon and another sterile neutrino of mass $m_2$ comparable to the mass of $m_1$. In this case, the term $hc/\lambda_e$ in Eq. \eqref{eq:I_lambda} can be replaced with $x\,m_1c^2$. Considering both $m_1$ and $m_2$ are of $\mathcal{O}(\rm keV)$, we can approximate $m_1+m_2\sim 2m_1$. The photon's energy produced from this process becomes $E_{\gamma} = m_1-m_2$. From hereinafter, we define $m_1-m_2 = \Delta m$. Consequently, the term $x$ mentioned earlier will be $\Delta m/m_1$. On substituting all the terms as mentioned above in Eq. \eqref{eq:I_lambda}, we rewrite the equation as
\begin{equation}
    I_\lambda = \frac{c}{4\pi}\frac{{\Omega_{\chi,0}}{\rho_c}{c^2}\Gamma_{m_1}}{\lambda_{obs}(1+z)H(z)}~\frac{\Delta m}{m_1}.
    \label{eq: ILambda}
\end{equation}
From the above equation, we can observe that $I_{\lambda}$ is directly proportional to the decay width and the energy of the released photon. However, it is inversely related to the mass of the decaying particle and decays over redshifts due to the expansion of the universe.
\newpage
\section{Results and Discussions}
\label{sec4}

In the previous section, we discussed that the intensity of photon production by the sterile neutrino radiative decays which is linked to the neutrino magnetic moment. Here, we investigate the reported COB anomalous intensities, that is $8.06\pm 1.92\,\rm nW/m^2/sr$ \cite{Lauer:2022fgc} and $2.99\pm 2.03~\,\rm nW/m^{2}/sr$ \cite{Postman:2024erl}, and analyse them within the framework of the neutrino magnetic moment to derive an upper bound on the transition magnetic moment. From hereinafter we will represent the reported intensities $8.06\pm 1.92\,\rm nW/m^2/sr$ and $2.99\pm 2.03~\,\rm nW/m^{2}/sr$ as $I_{806}$ \cite{Lauer:2022fgc} and $I_{299}$ \cite{Postman:2024erl}, unless stated otherwise. Specifically, we investigate the possible scenario: the sterile-to-sterile transition magnetic moment.

\subsection{Sterile to sterile neutrino decay}
\label{sec:sterile2sterile}

\begin{figure}
    \centering
    \subfigure[]{
        \includegraphics[width=\linewidth]{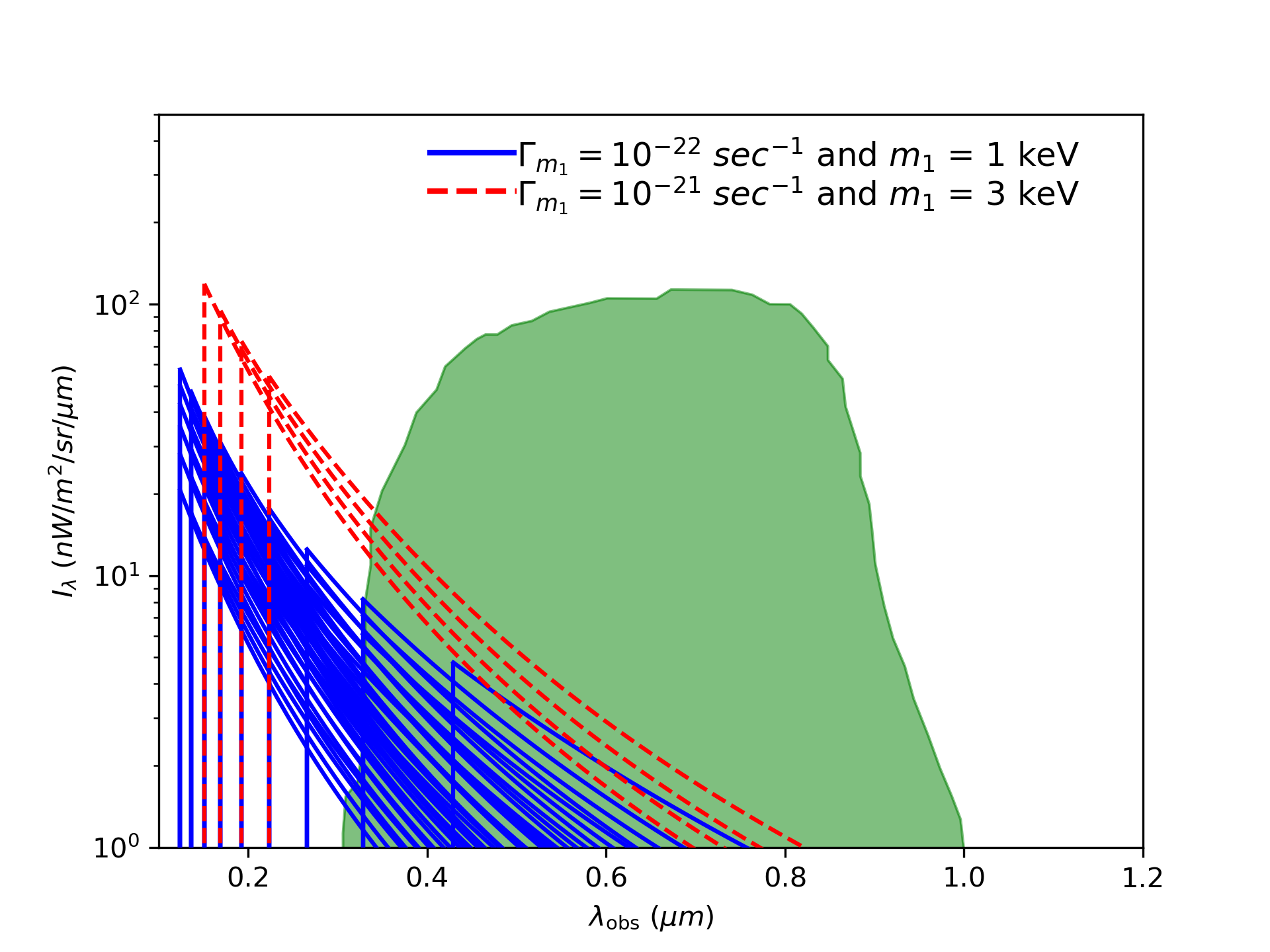}
        \label{Fig.2c}
    }
    \subfigure[]{
        \includegraphics[width=\linewidth]{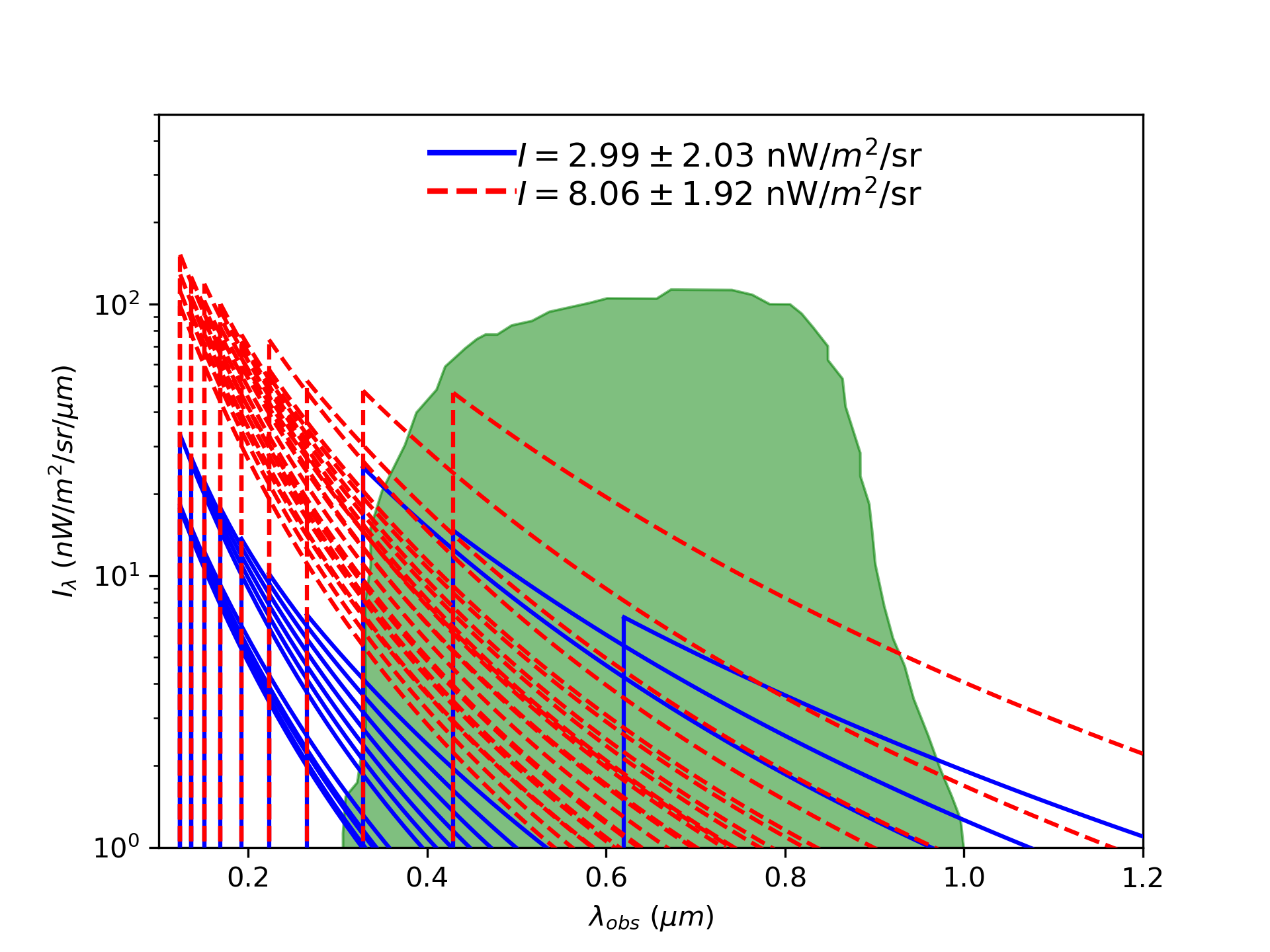}
        \label{Fig.2d}
    }
    \caption{(a) Illustrates the variation in specific intensity $I_{\lambda}$ with respect to $\lambda_{\rm obs}$, for $m_1 = 1\,\rm keV$ and $3\,\rm keV$ decaying at $\Gamma_{m_1} = 10^{-22} \, \text{sec}^{-1}$ and $\Gamma_{m_1} = 10^{-21} \, \text{sec}^{-1}$ respectively--- depicted in the blue solid and red dashed lines . (b) The variations in $I_{\lambda}$ contributing to LORRI's reported anomalous intensities. For $I = 2.99\pm 2.03\,\rm nW/m^2/sr$ \cite{Postman:2024erl}, we consider $1-40\,\rm keV$ sterile neutrinos decaying in the range $\Gamma_{m_1} \sim (10^{-22}-10^{-21}) \, \text{sec}^{-1}$ -- shown in blue solid lines. Further for $I = 8.06\pm 1.92~\,\rm nW/m^2/sr$ \cite{Lauer:2022fgc}, we consider $1-20\,\rm keV$ sterile neutrinos decaying in the range $\Gamma_{m_1} \sim (10^{-21}-10^{-20}) \, \text{sec}^{-1}$ -- shown in red dashed lines. The green-shaded area illustrates the shape of LORRI's responsivity for observed wavelengths \cite{bernal2022cosmic}. }
    \label{fig:keV_spec_intensity}
\end{figure}

\begin{figure}
    \centering
        \centering
        \includegraphics[width= \linewidth]{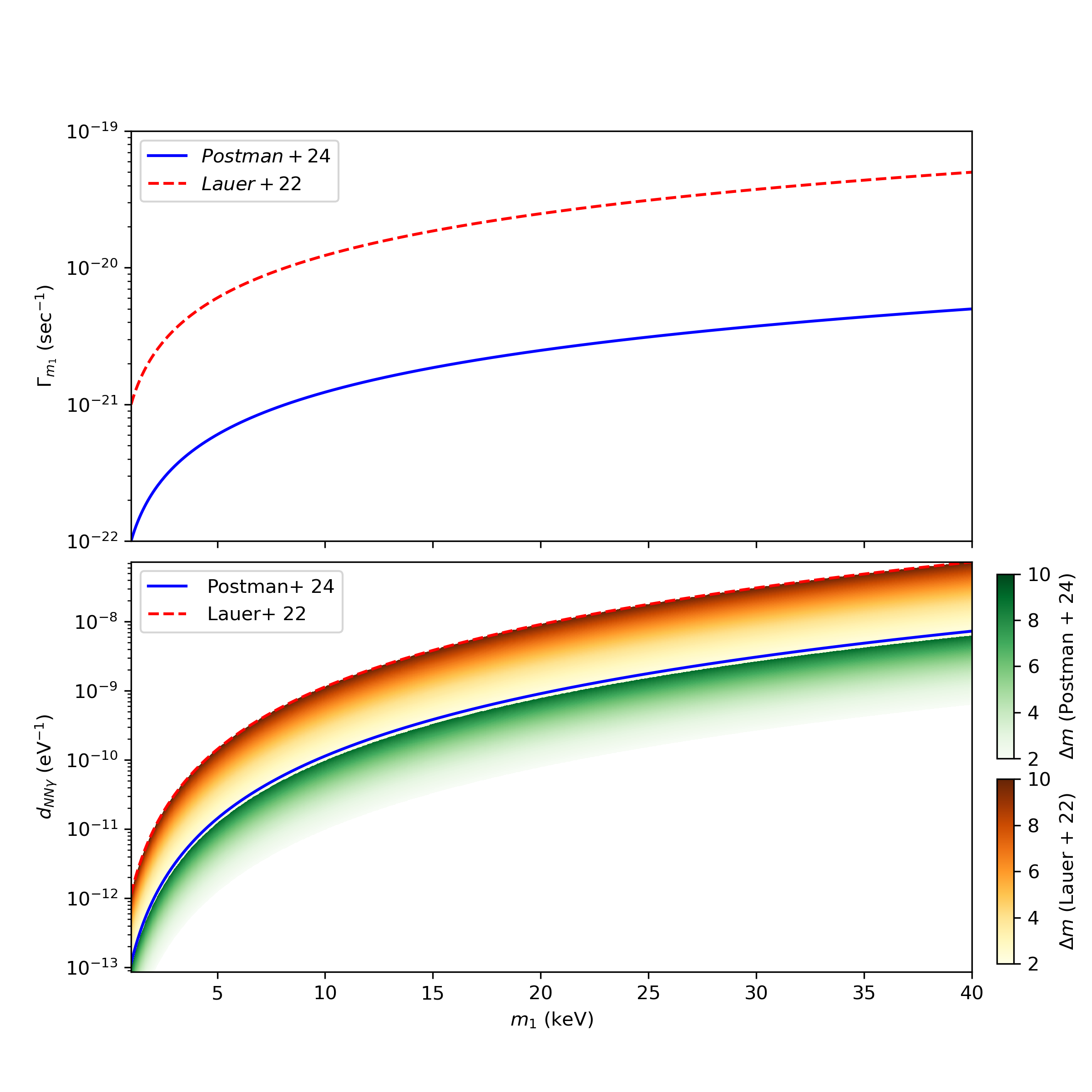}
        
        \caption{\emph{Top panel:} The minimum required decay width to obtain $I_{806}$ and $I_{299}$ shown in red dashed and blue solid lines, respectively, for masses $1 ~\rm keV - 40 ~\rm keV$. 
        \emph{Bottom panel:} The corresponding upper bounds on $d_{\rm NN\gamma}$ for the anomalous intensities $I_{806}$ \cite{Lauer:2022fgc} and $I_{299}$ \cite{Postman:2024erl}. The colour band represents the change of the sterile-to-sterile transition magnetic moment $(d_{NN\gamma})$ in $\,\rm eV^{-1}$ versus the mass of sterile neutrino $(m_{1})$ in keV scale, where, $\Delta m$ vary from $2-10\,\rm eV$. The green one represents the variation of $d_{NN\gamma}$ due to the anomalous rest intensity of COB, $2.99\pm 2.03\,\rm nW/m^{2}/sr$. Whereas, the yellow one represents the same variation of $d_{NN\gamma}$ due to the anomalous COB excess intensity $8.06\pm 1.92\,\rm nW/m^2/sr$. }
        \label{Fig.2}
\end{figure}

\begin{figure}
    \centering
    \includegraphics[width=1.1\linewidth]{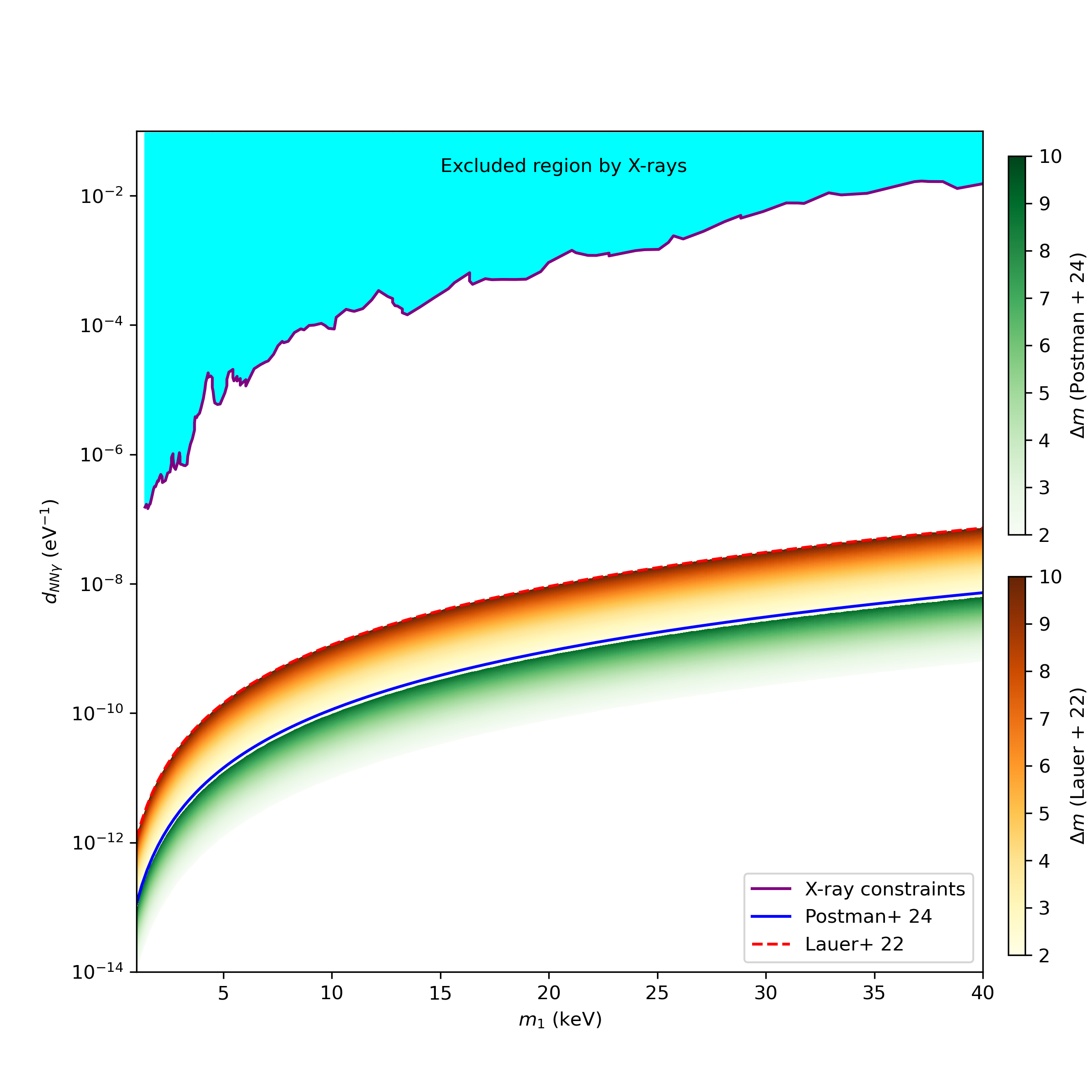}
    \caption{The yellow and green colour bands illustrate the variation of  $d_{\rm NN\gamma}$ versus the mass of the sterile neutrino $(m_1)$ with the changes of $\Delta m$, which vary from $2\,\rm eV$ to $10\,\rm eV$ . These two-colour bands denote the changes of $d_{NN\gamma}$ obtained from anomalous intensities of COB, $I_{806}$ \cite{Lauer:2022fgc} and $I_{299}$ \cite{Postman:2024erl}, respectively. The cyan-shaded region represents the excluded values of $d_{\rm NN\gamma}-m_1$ values from X-ray observations \cite{natwariya2022bounds}.}
    \label{fig:dnn_kev_plot}
\end{figure}

In this section, we calculate the required decay width and the transition magnetic moment for the anomalous intensities reported by LORRI \cite{Lauer:2022fgc, Postman:2024erl}.
We consider the keV mass range sterile neutrino as a dark matter candidate, which decays into another sterile neutrino and photon through the radiative decay process. We note that the sterile neutrinos with mass $< 0.4\,\rm keV$ are excluded as a dark matter candidate by a conservative application of the Tremaine-Gunn bound \cite{abazajian2006constraints}. Now, to calculate $I_\lambda$ required to explain the anomalous intensity of COB, we need to determine the energy of photons $(\Delta m)$, as shown in Eq. \eqref{eq: ILambda}. The LORRI observation reported flux of photons at pivot wavelength $0.608\,\mu m\,(\sim 2\,\rm eV)$ \cite{Lauer:2022fgc}. Therefore, $\Delta m$ should be in the range $2\,\rm eV-10\, \rm eV$, such that LORRI's band could observe these redshifting photons today. We restrict the energies of photons in this range, as the photons produced at $z\gtrsim 4$ will get absorbed by the intergalactic gas medium \cite{Madau:1995js, Inoue:2014zna}.

To illustrate the variation of $I_{\lambda}$ with $\lambda_{\rm obs}$, here we consider $1\,\rm keV$ and $3\,\rm keV$ masses of sterile neutrinos decaying at $10^{-22}\,\rm sec^{-1}$ and $10^{-21}\,\rm sec^{-1}$, respectively--- shown in blue solid lines and red dashed lines, respectively, in Fig. \ref{Fig.2c}. 
In this figure, the cutoff vertical line for wavelength $\lambda_e$ is the wavelength of photons observed today. 
The green-shaded region represents the shape responsivity of LORRI's camera with respect to the observed wavelength \cite{Weaver_2020, bernal2022cosmic, Porras-Bedmar:2024uql}. Furthermore, we adjust the intensities to match LORRI's reported values, as shown in Fig. \ref{Fig.2d}. We first fix the obtained intensity to $I_{806}$ \cite{Lauer:2022fgc}, for $1-20\, \rm keV$ sterile neutrinos decaying in the range $\Gamma_{m_1}\sim \left(10^{-21} - 10^{-20}\right)\, \rm sec^{-1}$, as depicted in the red dashed lines. Similarly, the blue solid lines show $1-40\,\rm keV$ sterile neutrinos decaying at $\Gamma_{m_1}\sim \left(10^{-22} - 10^{-21}\right)\, \rm sec^{-1}$ producing the LORRI's reported intensity $I_{299}$ \cite{Postman:2024erl}. We note that the intensity curves at lower $\lambda_{\rm obs}$ values suggest the production of photons with energy $\Delta m$ for the aforementioned $m_1$ and $\Gamma_{m_1}$ originating from higher redshifts. From Eq. \eqref{eq: ILambda}, we can observe that high energetic photons originating at higher redshifts require smaller $\Gamma_{m_1}$. Therefore, a sterile neutrino with a large decay width is required to produce less energetic photons in order to contribute to the anomalous COB intensities.  
In the next section, we present the minimum required $\Gamma_{m_1}$, and hence the $d_{\rm NN\gamma}$, to obtain the reported COB anomalous intensities for different $m_1$ values.

In the top panel of Fig. \eqref{Fig.2}, we plot the required minimum decay width for the mass range of sterile neutrinos $1\,\rm keV- 40\,\rm keV$ to obtain the anomalous intensities of COB. The red dashed and blue solid lines represent the minimum decay width for masses $1-40\,\rm keV$ to obtain the reported intensities in $I_{806}$ and $I_{299}$, respectively. We find that sterile neutrinos of mass $3\,\rm keV$ and $1\,\rm keV$ require $1\times 10^{-21}\,\rm sec ^{-1}$ and $8.54\times 10^{-22}\,\rm sec^{-1}$ for intensities reported in $I_{8.06\pm 1.92 }$ and $I_{2.99 \pm 2.03}$, respectively. The decay width of the sterile neutrino increases as we increase the mass. This can be analysed from the fact that $I_{\lambda}\propto\Gamma_{m_1}/m_1$--- shown in Eq. \eqref{eq: ILambda}. We restrict our analysis to $1\leq m_1/\rm keV\leq 40$ because a further increase in $m_1$ requires an increase in $\Gamma_{m_1}$. For instance, the required minimum decay width $\Gamma_{m_1}$ for $m_1 \gtrsim 40\,\rm keV$ requires $\gtrsim 10^{-20} \,\rm sec^{-1}$ for $I_{806}$, which becomes comparable to $\sim t^{-1}_{\rm U}$ which requires a consideration of variation in the dark matter density--- as mentioned in section \eqref{sec3}. We then plot the minimum values of $d_{\rm NN\gamma}$, in the bottom panel, for the above obtained $\Gamma_{m_1}$ values --- shown in the red dashed and blue solid lines. We find that under the conditions $m_1+m_2 \sim 2m_1$ and $m_1 - m_2 = \Delta m$, the decay width Eq. \ref{eq: dec width} becomes $16|d_{\rm NN\gamma}|^2\,\Delta m^3/\pi$. As mentioned earlier, $\Delta m$ takes values of $2\,\rm eV - 10\,\rm eV$ for the observed flux of photons. 
We find that the sterile-to-sterile transition magnetic moment $d_{\rm NN\gamma}/\rm eV^{-1}$ takes values of $(3\times 10^{-12}- 10^{-8})$ and $(3\times 10^{-13}- 10^{-9})$ for mass $1\,\rm keV -40\,\rm keV$--- as shown in the red dashed and blue solid lines, respectively. In Fig.~\ref{Fig.2}, we illustrate the colour band of $\Delta m$ variation from $2\,\rm eV$ to $10\,\rm eV$ for the two anomalous intensities of COB. This colour band represents the changes of the sterile-to-sterile transition magnetic moment $(d_{NN\gamma})$ in $\,\rm eV^{-1}$ versus the mass of the sterile neutrino $(m_1)$ for the changes of the minimum required decay width of the sterile neutrino. The green colour band represents the changes of the sterile-to-sterile transition magnetic moment $(d_{NN\gamma})$ for the anomalous rest intensity of COB, $2.99\pm 2.03\,\rm nW/m^{2}/sr$. Whereas, the yellow colour band region represents the changes $d_{NN\gamma}$ for the anomalous COB excess intensity $8.06\pm 1.92\,\rm nW/m^2/sr$. From the decay width Eq.~\eqref{eq: decay width}, we can see that the sterile-to-sterile transition magnetic moment $(d_{NN\gamma})$ depends directly on the decay width of the sterile neutrino. We observe that for the smaller mass difference, $\Delta m = 2~\,\rm eV$, the sterile-to-sterile transition magnetic moment $(d_{NN\gamma})$ approaches up to $10^{-13}\,\rm eV^{-1}$ and the minimum required decay width $(\Gamma_{m_1})$ reaches up to $10^{-21}\,\rm sec^{-1}$ for the anomalous COB excess intensity $8.06\pm 1.92 \,\rm nW/m^2/sr$ -- as shown in the red dashed line. Whereas, for the larger mass difference between two heavy sterile states, $\Delta m = 10~\,\rm eV$, the minimum required decay width of the sterile neutrino $(\Gamma_{m_1})$ increases and reaches up to $\sim 10^{-19}\,\rm sec^{-1}$ for the mass range of the sterile neutrino $m_1\geq 40\,\rm keV$. Similarly, the sterile-to-sterile transition magnetic moment $(d_{NN\gamma})$ also increases and approaches up to $10^{-8}\,\rm eV^{-1}$. So, we observe that for the mass range of the sterile neutrino $m_1\geq 40\,\rm keV$ or the larger mass difference between two heavy sterile states $\Delta m = 10\,\rm eV$, we obtain that the minimum required decay width of the sterile neutrino approaches up to $t_{\,\rm U}\sim 10^{18}\,\rm sec$. We illustrate the changes $d_{NN\gamma}$ due to the changes of mass difference $\Delta m $ from $ 2\,\rm eV$ to $10\,\rm eV$ in this Fig.~\ref{Fig.2}. Further in Fig. \eqref{fig:dnn_kev_plot}, we present the upper bound on the $d_{\rm NN\gamma}$ in the presence of existing X-ray bounds on $\Gamma_{m_1}-m_1$ values as the cyan-coloured region \cite{natwariya2022bounds}. For instance, the upper bound on the sterile-to-sterile transition magnetic moment $d_{NN\gamma}$ from X-ray observations is approximately $10^{-6}\,\rm eV^{-1}$ for the mass of sterile neutrino $5\,\rm keV$. In contrast, for the larger value of sterile neutrino mass, $40\,\rm keV$, the upper constraint on $(d_{NN\gamma})$ yields approximately $10^{-2}\,\rm eV^{-1}$ .

\subsection{Sterile to active neutrino decay}
In this section, we examine light sterile neutrinos in the mass range \( m_s \in [4,\,20]\,\mathrm{eV} \). These neutrinos behave as \textit{dark radiation} rather than cold dark matter. Consequently, their abundance is typically computed using the Dodelson–Widrow (DW) mechanism, given by \( \Omega_s h^2 = \frac{m_s}{94.1\,\mathrm{eV}} \). However, for sterile neutrino masses within this range, the DW mechanism is ruled out by cosmological observations, including data from the Cosmic Microwave Background (CMB) and large-scale structure formation~\cite{xu2022cosmological}.

Assuming a conservative minimum relic temperature of \( T_X = 0.91\,\mathrm{K} \), stringent cosmological constraints exist on the mass of thermal light relics. Ref.~\cite{xu2022cosmological} analyzes four categories of such relics—real scalars, Weyl fermions, Dirac fermions, and vectors. Their analysis shows that minimally coupled Weyl fermions with \( m_X > 2.3\,\mathrm{eV} \), real scalars with \( m_X > 11\,\mathrm{eV} \), and Dirac fermions with \( m_X > 1.1\,\mathrm{eV} \) are excluded at the 95\% confidence level~\cite{aker2019improved}. These bounds are notably more stringent—by factors of 2 to 5—than those reported in previous literature~\cite{xu2022cosmological}.

Furthermore, results from CMB analyses, weak gravitational lensing~\cite{abbott2022dark}, and galaxy surveys strongly constrain light thermal relics with masses \( m_X \gtrsim 0.1\,\mathrm{eV} \). Therefore, the observed excess in the Cosmic Optical Background (COB) intensity cannot be attributed to the decay of sterile neutrinos into active neutrinos within the considered mass range of \( 4\,\mathrm{eV} \leq m_s \leq 20\,\mathrm{eV} \).

\subsection{Active-to-active neutrino transition magnetic moment}

In the UV model, the active-to-active neutrino transition magnetic moment is naturally suppressed \cite{beltran2024probing}. Active-sterile mixing can create an active-to-active neutrino magnetic moment in the broken phase parameters when the right-handed neutrinos are integrated out \cite{beltran2024probing}. As such $d_{\nu\nu\gamma}\lesssim 2\times 10^{-8}~\text{\rm GeV}^{-1}$, we may anticipate that the active-to-active neutrino magnetic moments can satisfy the constraints from TEXONO \cite{li2003limit}, GEMMA \cite{beda2010gemma}, LSND, and Borexino \cite{agostini2017limiting} in the inverse see-saw mechanism, where larger active-sterile mixing can occur. For the specified mass range $(0.12~ \text{\rm eV} - 0.7~\text{\rm eV})$ of active neutrinos, we are not able to obtain the specific intensity of photons due to the radiative decay of active neutrinos. As for this aforementioned mass range of active neutrinos, a very less number of photons can be generated, which cannot explain the COB anomalous intensity in the observed wavelength range. For that reason, we cannot observe any specific intensity of photons in the observed wavelength range $\lambda_{obs} = 0.2 ~\mu m - 1.2 ~ \mu m$ because to explain the COB anomaly, we need minimum $4$~\rm eV mass of sterile neutrinos. Hence, the COB anomalous intensities can not be accounted for by the decay of active neutrinos into active neutrinos and photons.

\section{UV Origin of Small Mass Splittings in keV Sterile Neutrinos}
\label{sec:uv}

Although our primary analysis adopts an effective field theory approach, a complete understanding of the small eV-scale mass splitting between keV sterile neutrinos benefits from a consistent ultraviolet (UV) completion. Several well-motivated frameworks exist that naturally explain the emergence of quasi-degenerate sterile neutrino masses without fine-tuning. Two particularly appealing UV approaches are as follows:

\begin{enumerate}
    \item \textbf{Froggatt–Nielsen Mechanism:}
In models with a $U(1)_{\rm FN}$ Froggatt–Nielsen symmetry~\cite{Froggatt:1978nt,Pilaftsis:2003gt}, sterile neutrinos are assigned flavor-dependent charges under $U(1)_{\rm FN}$, and their Majorana masses arise from non-renormalizable operators suppressed by a high scale $\Lambda_{\rm FN}$ (the Froggatt–Nielsen (FN) cutoff scale, typically taken to be the GUT or Planck scale). The FN symmetry constrains the structure of the mass matrix, enforcing degeneracy at leading order. When the scalar field $S$ acquires a vacuum expectation value (VEV), the symmetry is spontaneously broken, generating mass terms suppressed by powers of the small parameter $\epsilon = \langle S \rangle / \Lambda_{\rm FN}$. Subleading contributions to the mass splitting may also arise from higher-order corrections or small perturbations consistent with the FN charge assignments. These effects induce naturally small mass splittings of the form
\[
\Delta m \sim \epsilon^n \Lambda_{\rm FN}
\]
where $n$ depends on the FN charge assignment and operator dimension.

In particular, if two sterile neutrinos $N_1$ and $N_2$ are assigned suitable FN charges, they can acquire masses at the keV scale via higher-dimensional operators suppressed by $\epsilon = \langle S \rangle / M_{\rm GUT}$. The resulting Majorana mass terms exhibit the structure:
\[
m_{N_{1,2}} \sim \epsilon M \sim \text{keV}, \quad \Delta m_{12} \sim \epsilon^2 M \sim \text{eV}
\]
for $M \sim \text{TeV}$, where 
$M$ denotes the mass of a heavy mediator fermion $X$ that appears in the UV completion and is integrated out to generate the effective operator and $\epsilon \sim 10^{-6}$. This setup yields quasi-degenerate keV-scale sterile neutrinos with eV-level splittings—without requiring fine-tuning—making them viable dark matter candidates. The underlying process is diagrammatically shown in Figure~\ref{fig:feynman_diag}.

\begin{figure}[h]
    \centering
    \begin{tikzpicture}
        \begin{feynman}
            \vertex (a) at (-3, 0) {\(N^c\)};
            \vertex (b) at (3, 0) {\(N\)};
            \vertex (v1) at (-1, 0);
            \vertex (v2) at (1, 0);
            \vertex (s1) at (-1, 1.2) {\(\langle S \rangle\)};
            \vertex (s2) at (1, 1.2) {\(\langle S \rangle\)};
            \node at (0, -0.7) {\(\varepsilon^2 M\)};

            \diagram* {
                (a) -- [fermion] (v1) -- [fermion, edge label=\(X\)] (v2) -- [fermion] (b),
                (s1) -- [scalar] (v1),
                (s2) -- [scalar] (v2),
            };
        \end{feynman}
    \end{tikzpicture}
    \caption{Majorana mass generation for sterile neutrino \(N\) via two insertions of the FN scalar \(S\), whose VEV breaks the \(U(1)_{\rm FN}\) symmetry. The interaction is mediated by a heavy fermion \(X\) of mass \(M\).}
    \label{fig:feynman_diag}
\end{figure}

    \item \textbf{Clockwork or Radiative Symmetry Breaking:}
    In alternative frameworks, the degeneracy of sterile neutrino masses is protected by a discrete or continuous symmetry that is broken radiatively. This results in small loop-induced splittings between otherwise degenerate states. The clockwork mechanism, in particular, achieves exponential suppression of mass terms through a chain of fields with nearest-neighbor interactions \cite{Giudice:2016yja,Maharana:2023wyq,Hong:2019bki,Hambye:2004jf, Blanchet:2009bu}. When applied to sterile neutrinos, such setups can yield keV-scale mass eigenstates with eV-level splittings, making them phenomenologically viable and technically natural.
\end{enumerate}

These UV completions demonstrate that small splittings among sterile neutrinos can emerge from structural features of the theory, rather than ad hoc fine-tuning, and are compatible with broader frameworks like seesaw and extra-dimensional models. Thus, the small $\mathcal{O}(\text{eV})$ mass difference between keV sterile states—essential for oscillation and decay phenomenology—can be realized in a variety of UV-consistent ways.

\section{Conclusion and Outlook}
\label{sec:Conclusion}

Recent measurements of the cosmic optical background (COB) photon flux intensity by LORRI have revealed some excess intensity with no known sources. This excess may indicate a potential source of energetic photons beyond the Standard Model (BSM). In this study, we investigated the possibility of sterile neutrinos decaying into photons and explored the corresponding decay widths. Notably, we did not account for photon energy generation through cascades involving other decay products. We found the required decay width of the sterile neutrino and obtained upper bounds on the sterile-to-sterile transition magnetic moments of the order $\mathcal{O}(10^{-22}-10^{-21})~\text{sec}^{-1}$ for the keV mass range of the decaying sterile neutrino.
Here, we adopt a minimal framework involving sterile neutrinos using the effective field theory approach. The radiative decay of sterile neutrinos, resulting in photon production, is primarily governed by magnetic moments associated with the sterile-to-sterile 
transition magnetic moment $(d_{NN\gamma})$. We calculate the $d_{NN\gamma}$ for the different $\Delta m$ values, which vary from $2-10\,\rm eV$. We find that, $d_{NN\gamma}$ take values of $3\times 10^{-12}\,\rm eV^{-1} -10^{-9}\,\rm eV^{-1}$ for sterile neutrinos of mass range $1\,\rm keV-40\,\rm keV$. In our work, we explain the bounds on sterile-to-sterile transition magnetic moments by considering the two COB measurements: one is $2.99\pm 2.03\,\rm nW/m^2/sr$ intensity, and the other one is due to anomalous COB excess intensity $8.06\pm 1.92\,\rm nW/m^2/sr$. Here, we have plotted the specific intensity for both COB measurements. Motivated by this work, one can probe CUB, CXB, and CIB measurements to constrain the sterile-to-sterile transition magnetic moment. We leave this as future work. Some upcoming experiments, such as GALEX and ULTRASAT satellites, can probe and constrain the decaying dark matter in the ultraviolet bandwidth \cite{Libanore:2024hmq}. Our minimal framework based on decaying sterile neutrinos of mass $\mathcal{O}$(keV) can contribute to the unknown excess COB intensity, thus deriving upper bounds on the sterile-to-sterile transition magnetic moments. In conclusion, our analysis establishes that the anomalous rest-frame intensity of the Cosmic Optical Background enables constraints on the sterile-to-sterile transition magnetic moment, \( d_{NN\gamma} \), while scenarios involving sterile-to-active or active-to-active transitions remain disfavored within this framework.

\section{Acknowledgement}
We sincerely acknowledge the usage of \texttt{python} packages \texttt{scipy}\footnote{\href{https://www.scipy.org/}{https://www.scipy.org/}} and \texttt{astropy}\footnote{\href{https://www.astropy.org/index.html}{https://www.astropy.org/index.html}}. We thank Rahul Kothari for his earlier collaboration on this project. H. H. acknowledges support from the Ministry of Higher Education, Government of India. T. S. is supported by the National Natural Science Foundation of China under Grants No. 12475094, No. 12135006, and No. 12075097.


%

\end{document}